%% file: main.tex
\documentclass[10pt,conference]{IEEEtran}
\IEEEoverridecommandlockouts
\usepackage{cite}

\usepackage{tcolorbox} 
\usepackage{csquotes}
\usepackage{amsmath,amssymb,amsfonts}
\usepackage{algorithmic}
\usepackage{graphicx}
\graphicspath{{./figs/}}
\DeclareGraphicsExtensions{.png,.eps,.pdf}
\usepackage{textcomp}
\usepackage{xcolor}
\usepackage{booktabs}
\usepackage{multirow}
\usepackage{enumerate}
\usepackage{xspace}

\usepackage{balance}
\usepackage[colorlinks,bookmarksopen,bookmarksnumbered,citecolor=red, urlcolor=black]{hyperref}
\def\BibTeX{{\rm B\kern-.05em{\sc i\kern-.025em b}\kern-.08em
    T\kern-.1667em\lower.7ex\hbox{E}\kern-.125emX}}
    
\usepackage{fancybox}
\newcommand{\hypobox}[1]{\begin{center}%
		\noindent\thicklines\setlength{\fboxsep}{8pt}%
		\cornersize{0.0}\Ovalbox{\begin{minipage}{3in}%
				#1\end{minipage}} \end{center}}

\usepackage{expl3}

\begin{document}

\newcommand{\etal}{\textit{et~al.~}}
\newcommand{\actfmt}[1]{#1}
\newcommand{\taskfmt}[1]{\textit{#1}}
\newcommand{\blog}[1]{[\texttt{#1}]}



\newcommand{\citep}[1]{\cite{#1}}
\newcommand{\addcitet}[2]{\csdef{mapcitet#1}{#2}}
\newcommand{\citet}[1]{\csuse{mapcitet#1}\cite{#1}}
\input{ieee_cite_map.tex}

\title{Software Engineering and Foundation Models: 
Insights from Industry Blogs Using a Jury of Foundation Models
}

\author{\IEEEauthorblockN{Hao Li}
\IEEEauthorblockA{\textit{Queen’s University} \\
Kingston, Canada \\
hao.li@queensu.ca}
\and
\IEEEauthorblockN{Cor-Paul Bezemer}
\IEEEauthorblockA{\textit{University of Alberta} \\
Edmonton, Canada \\
bezemer@ualberta.ca}
\and
\IEEEauthorblockN{Ahmed E. Hassan}
\IEEEauthorblockA{\textit{Queen’s University} \\
Kingston, Canada \\
ahmed@cs.queensu.ca}
}

\maketitle

\begin{abstract}
\input{sections/abstract}
\end{abstract}

\begin{IEEEkeywords}
Foundation models, FM4SE, SE4FM, LLM-as-a-judge, industry trends, LLM
\end{IEEEkeywords}

\newcommand{\rqone}{Which FM4SE activities are discussed in industry blog posts?}
\newcommand{\rqtwo}{Which SE4FM activities are discussed in industry blog posts?}
\newcommand{\rqthree}{What are the key challenges and approaches discussed by practitioners in SE4FM blogs?}

\newcommand{\ourmethod}{LLM Jury\xspace}

\newcommand{\motivation}{\noindent\textit{Motivation. }}
\newcommand{\approach}{\medskip\noindent\textit{Approach. }}
\newcommand{\findings}{\medskip\noindent\textit{Findings. }}

\input{sections/introduction}
\input{sections/background}

\begin{figure*}[t]
\centering
\includegraphics[width=\textwidth]{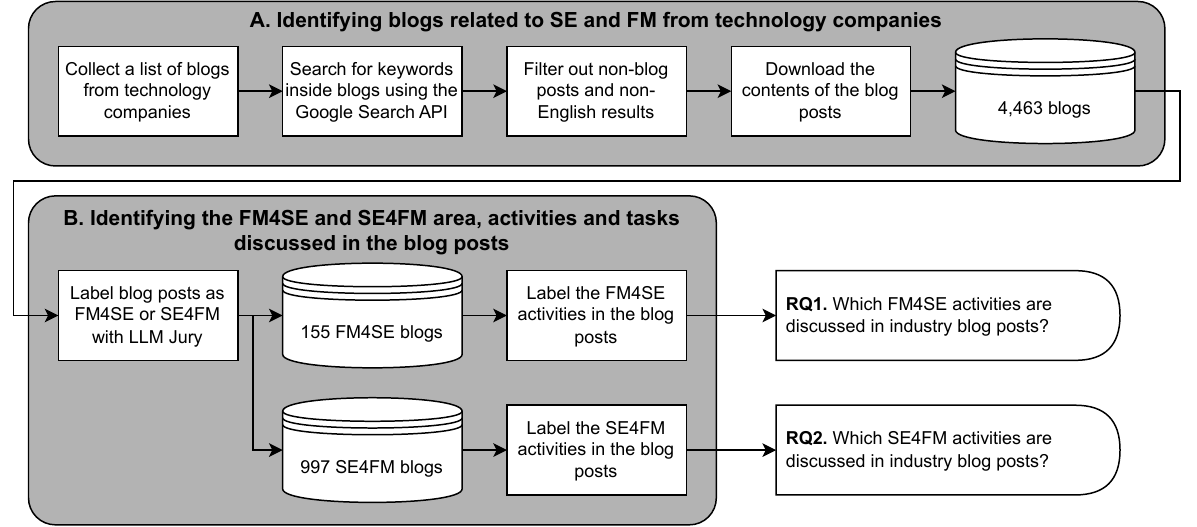}
\caption{An overview of our methodology.}
\label{fig:methodology}
\end{figure*}

\input{sections/llm_jury}
\input{sections/methodology}
\input{sections/results}

\input{sections/discussion}
\input{sections/threats}
\input{sections/conclusion}

\balance
\bibliographystyle{IEEEtranS}
\bibliography{IEEEabrv, mybib}

\end{document}

%% file: ieee_cite_map.tex
\addcitet{llm4se_survey_2024}{Hou et~al.}
\addcitet{se4ai_survey_2022}{Mart\'{\i}nez-Fern\'{a}ndez et~al.}
\addcitet{bangash2019ml_so}{Bangash et~al.}
\addcitet{Morovati2023}{Morovati et~al.}
\addcitet{butler2012think}{Butler}
\addcitet{zhang_judge_2024}{Zheng et~al.}
\addcitet{verga_juries_2024}{Verga et~al.}
\addcitet{kocmi_llmjudge_2023}{Kocmi and Federmann}
\addcitet{ouyang_rlhf_2022}{Ouyang et~al.}
\addcitet{panickssery_llm_selffavor_2024}{Panickssery et~al.}
\addcitet{wang_llm4test_2024}{Wang et~al.}
\addcitet{fan_llm4se_2023}{Fan et~al.}
\addcitet{chang_se4llm_evaluation_2024}{Chang et~al.}
\addcitet{villamizar_se4ml_req_2021}{Villamizar et~al.}
\addcitet{masuda_se4ml_quality_2018}{Masuda et~al.}
\addcitet{liu_prompteng_2023}{Liu et~al.}
\addcitet{wei_cot_2024}{Wei et~al.}
\addcitet{brown_fewshot_2020}{Brown et~al.}
\addcitet{lin2022teaching}{Lin et~al.}
\addcitet{mielke_reducing_2022}{Mielke et~al.}
\addcitet{xiong2024can}{Xiong et~al.}
\addcitet{emam_kappa_1999}{Emam}
\addcitet{chiang_chatbotarena_2024}{Chiang et~al.}
\addcitet{yang_qwen2_2024}{Yang et~al.}
\addcitet{replication_package}{Li et~al.}
\addcitet{amershi_se4ml_2019}{Amershi et~al.}
\addcitet{hassan_aiware_2024}{Hassan et~al.}
\addcitet{fried2023incoder}{Fried et~al.}
\addcitet{dubey2024llama3herdmodels}{Dubey et~al.}
\addcitet{hf_llama3}{Schmid et~al.}
\addcitet{chen2023acceleratinglargelanguagemodel}{Chen et~al.}
\addcitet{zhu_modelcompress_2024}{Zhu et~al.}
\addcitet{rombach2021highresolution}{Rombach et~al.}
\addcitet{lewis_rag_2020}{Lewis et~al.}
\addcitet{zhang2023recognizeanythingstrongimage}{Zhang et~al.}
\addcitet{pan_codetrans_2024}{Pan et~al.}
\addcitet{liang_codeassist_2024}{Liang et~al.}
\addcitet{jiang_loadtest_2015}{Jiang and Hassan}
\addcitet{wei2022emergentabilitieslargelanguage}{Wei et~al.}
\addcitet{hassan_se3_2024}{Hassan et~al.}

%% file: sections/abstract.tex
Foundation models (FMs) such as large language models (LLMs) have significantly impacted many fields, including software engineering (SE). The interaction between SE and FMs has led to the integration of FMs into SE practices (FM4SE) and the application of SE methodologies to FMs (SE4FM). While several literature surveys exist on academic contributions to these trends, we are the first to provide a practitioner's view. We analyze 155 FM4SE and 997 SE4FM blog posts from leading technology companies, leveraging an FM-powered surveying approach to systematically label and summarize the discussed activities and tasks. 
We observed that while code generation is the most prominent FM4SE task, FMs are leveraged for many other SE activities such as code understanding, summarization, and API recommendation. The majority of blog posts on SE4FM are about model deployment \& operation, and system architecture \& orchestration. Although the emphasis is on cloud deployments, there is a growing interest in compressing FMs and deploying them on smaller devices such as edge or mobile devices.
We outline eight future research directions inspired by our gained insights, aiming to bridge the gap between academic findings and real-world applications. Our study not only enriches the body of knowledge on practical applications of FM4SE and SE4FM but also demonstrates the utility of FMs as a powerful and efficient approach in conducting literature surveys within technical and grey literature domains. Our dataset, results, code and used prompts can be found in our online replication package at \url{https://zenodo.org/records/14563992}.

%% file: sections/introduction.tex
\section{Introduction}\label{sec:introduction}

In recent years, the rapid advancements in machine learning (ML) have fundamentally transformed various fields, including software engineering (SE). Among these developments, foundation models (FMs) such as large language models (LLMs) have emerged as a major force, reshaping how software is developed, tested, and maintained~\citep{llm4se_survey_2024}. The interaction between SE and FMs has led to the emergence of two key trends: (1)~FMs for SE~(FM4SE), where FMs are leveraged to automate or enhance various SE tasks, such as code generation and testing, and (2)~SE for FMs~(SE4FM), where SE practices are adapted to the development and deployment of FMs.

Academic research has made significant strides in exploring these trends, but literature surveys have only focused on published, peer-reviewed literature~\citep{llm4se_survey_2024, se4ai_survey_2022}, mostly leaving out the perspectives and experiences of industry practitioners. The input of practitioners, who work at the intersection of SE and FMs in real-world settings, is a crucial yet underexplored source of insights. While the research community has recognized the value of user-generated contents such as Q\&A websites~\citep{bangash2019ml_so} and issue reports~\citep{Morovati2023},  less attention has been paid to grey literature such as technical blog posts from industry leaders. Tech companies publish blog posts for several reasons, including positioning themselves as innovation leaders and establishing thought leadership~\cite{butler2012think}. As a result, these blog posts often provide in-depth discussions on cutting-edge challenges and solutions in SE and FM integration. 

\begin{figure}[t]
\centering
\includegraphics[width=\linewidth]{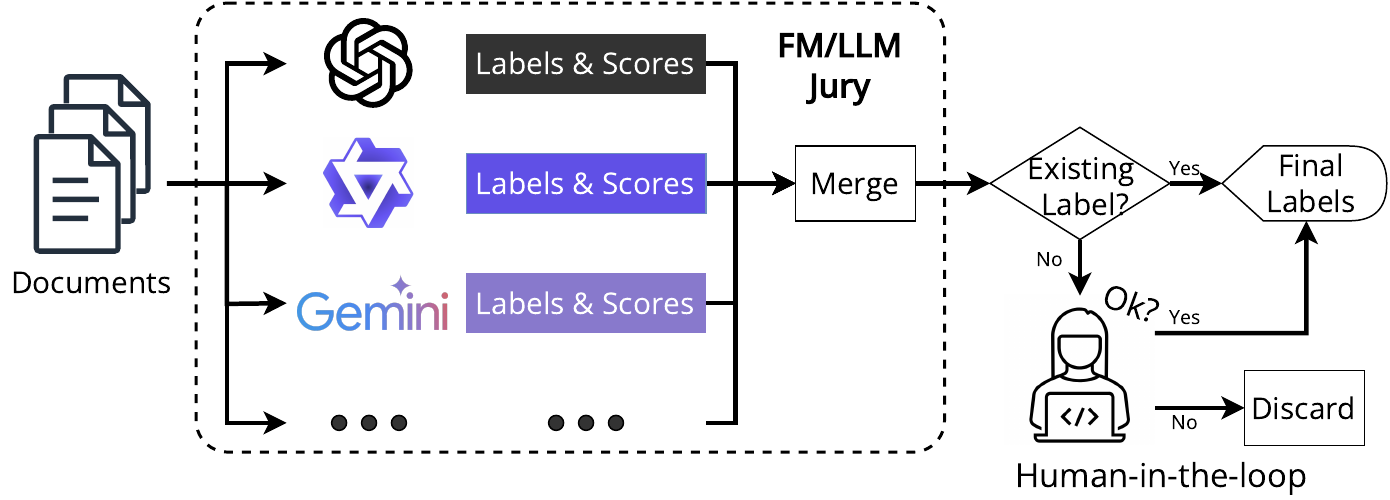}
\caption{An overview of our FM/LLM Jury approach for labelling blog posts. A blog post is labelled by every FM in the jury, and the final label is selected through a majority vote (using the normalized confidence score as a tie-breaker).}
\label{fig:blog_llm_juries}
\end{figure}


To bridge the gap between academic findings and industry practices, we analyze blog posts from leading technology companies, focusing on how practitioners discuss the challenges and approaches related to FM4SE and SE4FM. By systematically labelling and examining these blog posts, we seek to provide a clearer picture of how FMs are being integrated into the SE domain (i.e., FM4SE), and how SE principles are being applied to FMs (i.e., SE4FM) in industry. This study stands out by offering a synthesized industry voice, derived directly from real-world, practitioner-driven insights. We employ an ensemble of FMs as judges~\citep{zhang_judge_2024} into an FM/LLM Jury~\citep{verga_juries_2024} (see Figure~\ref{fig:blog_llm_juries}) to assist with the labelling and synthesis of knowledge within 155 FM4SE and 997 SE4FM blog posts. Our study focuses on these research questions (RQs):

\begin{enumerate}[\bfseries RQ1.]
\item \textbf{\rqone} 
Software development tasks, particularly code generation, are the most frequently discussed across FM4SE blogs. FMs are increasingly integrated as code assistants, providing developers with multifunctional tools to boost productivity. Vulnerability detection is the dominant software quality assurance task, while software maintenance activities primarily focus on refactoring and transforming existing codebases.

\item \textbf{\rqtwo}
The most discussed activities in SE4FM blog posts are model deployment \& operation, with a focus on cloud hosting and model serving \& scaling. Other trends include prompt chaining, workflow orchestration, and building AI agents. Data management activities focus on RAG and vector databases to support unstructured data and information retrieval. Model customization relies on fine-tuning methods such as full fine-tuning, LoRA, and RLHF. 



\end{enumerate}

The main contributions of this paper are:
\begin{itemize}
    \item The first study of industry blog posts on FM4SE and SE4FM to provide the practitioner's view on these emerging and crucial topics in today's software industry.
    \item A dataset of 1,152 blog posts from top technology companies related to FM4SE or SE4FM. 
    \item A list of eight research directions that are driven by the findings of our survey on blog posts.
    \item A demonstration of an efficient approach that leverages a jury of FMs to assist with grey literature surveys on SE-related topics.
\end{itemize}

\textbf{Paper Organization.} The rest of this paper is organized as follows. Section~\ref{sec:background} presents background information and related work. Section~\ref{sec:llm_jury} details the proposed FM/LLM Jury framework. Section~\ref{sec:methodology} presents our methodology. Sections~\ref{sec:rq1} and~\ref{sec:rq2} present the findings of our research questions. Section~\ref{sec:discussion} discusses promising future research directions that follow from our survey. Section~\ref{sec:threats} discusses the threats to the validity our study. Section~\ref{sec:conclusion} concludes this paper.

%% file: sections/background.tex
\section{Background and Related Work}\label{sec:background}

\subsection{LLM-as-a-judge}

Leveraging FMs/LLMs as evaluators, or LLM-as-a-judge, has emerged as a scalable alternative to traditional human evaluations to assess the quality of outputs from LLMs~\citep{kocmi_llmjudge_2023}. This assessment is not straightforward, as LLMs generate natural language, which needs to be compared with a ground truth semantically. LLM-as-a-judge builds on the idea that state-of-the-art models, especially those trained with Reinforcement Learning from Human Feedback (RLHF)~\citep{ouyang_rlhf_2022} (e.g., GPT-4) are well-aligned with human judgments, making them promising substitutes~\citep{zhang_judge_2024}. While the use of FMs like GPT-4 as evaluators has become more common, these models often exhibit biases, such as favouring their own outputs over those from other models~\citep{panickssery_llm_selffavor_2024}. To mitigate these biases, researchers have proposed the use of a panel of FM evaluators instead of relying on a single model \citep{verga_juries_2024}. Instead of using a single FM/LLM to evaluate FM/LLM outputs, in this paper, we propose using a jury of FMs/LLMs to assist with the labelling and summarization of industry blog posts.

\subsection{Related work}
The work that is closest related to our work consists of other literature surveys on FM4SE and SE4FM. 
Recent comprehensive surveys on FM4SE have examined the rapidly growing field of FMs/LLMs applied to SE activities and tasks. \citet{llm4se_survey_2024} conducted a systematic survey of 395 studies covering the application of LLMs to 84 specific SE tasks across 6 SE activities. In addition, \citet{wang_llm4test_2024} surveyed 102 studies that have used LLMs for software testing. These applications have shown promise, but several challenges remain. For example, \citet{fan_llm4se_2023} emphasized technical challenges like hallucinations when applying LLMs for SE, and highlighted the importance of hybrid techniques that combine traditional SE with LLMs.

Other surveys (on SE4FM) focus on how established SE practices can be adapted to support building, testing, deploying and maintaining FMs. \citet{chang_se4llm_evaluation_2024} reviewed evaluation methods and benchmarks for LLMs in different areas such as education and social sciences, highlighting the importance of robust benchmarks to assess the performance of LLMs. Most prior surveys on SE for models have focused on SE4ML rather than SE4FM (i.e., SE for machine learning models that are not foundation models). \citet{se4ai_survey_2022} reviewed 248 studies and classified them based on the Software Engineering Body of Knowledge (SWEBOK), identifying gaps in areas like maintenance and data handling. \citet{villamizar_se4ml_req_2021} discussed gaps in requirements engineering for ML, while \citet{masuda_se4ml_quality_2018} surveyed quality assurance approaches, highlighting the need for specialized testing techniques in verifying an ML system's output.

All prior surveys on FM4SE and SE4FM focused on academic efforts. Our work is the first to provide an overview of FM4SE and SE4FM activities in top technology companies.



%% file: sections/llm_jury.tex
\section{Using an FM/LLM Jury for Labelling Blog Posts}\label{sec:llm_jury}

Labelling blog posts using a single frontier FM (e.g., GPT-4o) in the LLM-as-a-judge approach~\citep{zhang_judge_2024} can be both expensive and potentially biased. To address these limitations, we propose FM/LLM Jury, a methodology that leverages multiple FMs to collaboratively label blog posts. In this framework, each model provides a label along with a confidence score, and these outputs are merged using a majority vote to determine a final label. This framework is inspired by \citet{verga_juries_2024}. 

\subsection{Constructing the prompt}

The prompt construction process is iterative and consists of the following key steps:

\textit{Step 1 -- Create the golden dataset.} 
To evaluate the performance of the prompts that we send to the FM and the quality of our FM/LLM Jury, we begin by constructing a golden dataset. We randomly sample a subset of blog posts and manually label them to serve as ground truth. 

\textit{Step 2 -- Design the prompt.} 
We design prompts following best practices in prompt engineering~\cite{openai_prompt} and techniques outlined by~\citet{liu_prompteng_2023}. The prompt should instruct the FMs on how to label the blog posts, specifying the labelling criteria and providing the necessary context. To improve labelling accuracy, we incorporate advanced techniques such as Chain-of-Thought prompting~\citep{wei_cot_2024} and few-shot in-context learning~\citep{brown_fewshot_2020}. We used a predefined set of labels to ensure a common vocabulary for the classification. Because FMs generate natural language text, a common vocabulary is necessary to (1)~group many different blog posts that may use different terms to describe the same aspects and (2)~facilitate a comparison with prior work which also uses that vocabulary. 
We also ask FMs to provide new labels if the predefined ones do not fit the content. Our replication package~\citep{replication_package} includes all used prompts. We encourage researchers to refine and rerun the process with new blog posts to ensure relevance in this rapidly moving field.

\textit{Step 3 -- Run the prompt on the golden dataset.} 
Each FM in the jury is prompted to label the blog posts in the golden dataset. For each blog post, the FM outputs both a label and an associated confidence score. As FMs may exhibit overconfidence, directly using the raw confidence scores is likely to introduce bias~\citep{lin2022teaching}. While calibration methods exist to address this issue, they typically require access to the model's internal information or fine-tuning~\citep{mielke_reducing_2022}, which is not feasible with closed-source FMs. Therefore, we apply a z-score standardization based on the confidence score distribution across the dataset to normalize the confidence values.

\textit{Step 4 -- Compare FM labels with human labels.} 
The produced labels are compared with the human-provided labels from the golden dataset. We assess inter-rater reliability using Cohen's $\kappa$ coefficient~\citep{emam_kappa_1999}, which measures the degree of agreement between the FM-generated labels and human labels. We set a threshold of $\kappa > 0.78$ (indicating excellent agreement) for at least one FM in the jury. Additionally, all FMs must achieve $\kappa \geq 0.63$ (indicating substantial agreement). If these thresholds are not met, we return to Step 2 and refine the prompt iteratively. The refinement process involves adding clarifications, improving instructions, or reordering prompt components to resolve ambiguities~\citep{googlePromptIteration}. The iterative loop continues until the desired agreement is reached.

\textit{Step 5 -- Freeze the prompt for full dataset labelling.} 
Once the prompt achieves the required level of agreement in Step 4, it is finalized and used to label the entire dataset.

\subsection{Merging the FM outputs}

After the individual FMs in the jury provide their labels, we aggregate the results using a majority vote, where the final label is determined by the label that receives the most votes from the individual FMs. 
In case of a tie, we use the normalized confidence scores to break the tie. 

\subsection{Human-in-the-loop}
A human-in-the-loop process was employed to decide whether to accept or reject new labels proposed by FMs. However, the process is only required when the FM/LLM Jury cannot resolve the final label. In our study, the FM/LLM Jury successfully handled all cases without the need for human involvement in the labelling process.


\subsection{Selected FMs}
The jury FMs are selected based on their performance and ability to follow complex instructions. We used the LLM Arena Leaderboard~\citep{chiang_chatbotarena_2024} to identify strong candidates. We selected the open-source Qwen2-72B-Instruct~\citep{yang_qwen2_2024} model, and two closed-source models, GPT-4o-mini~\citep{openai_gpt4o_mini} and Gemini-1.5-Flash~\citep{geminiteam_2024}. These models were selected due to the balance between cost and performance. We did not select the top-performing models, as they are around ten times as expensive. 

%% file: sections/methodology.tex
\section{Methodology}\label{sec:methodology}

Figure~\ref{fig:methodology} gives an overview of our methodology. In this section, we discuss every step.

\subsection{Identifying blogs related to SE and FM from technology companies}

We employed a systematic approach using search queries based on keywords related to FM4SE and SE4FM to gather blog posts. The data collection process consists of these steps:

\input{tables/company_blog_sites}
\textit{Step 1 -- Collect a list of blogs from technology companies.} We began by collecting a list of companies from the \enquote{Technology} sector with a market capitalization greater than \$200 billion (\enquote{Mega}) based on data from NASDAQ.\footnote{\url{https://www.nasdaq.com}} In addition, we manually went through the top 100 companies from Forbes' Global 2,000 list~\citep{forbes} and included companies that are categorized under the \enquote{IT Software \& Services} industry. In total, we included 20 companies. For each of these companies, we searched for relevant blogs using this query:

\smallskip
\noindent\fbox{%
\parbox{\dimexpr\linewidth-2\fboxsep-2\fboxrule}{%
blog AND (\enquote{software} OR \enquote{research} OR \enquote{engineer} OR \enquote{engineering})
}%
}
\smallskip

We identified 35 blogs from 17 companies (see Table~\ref{tab:company_blog_sites}). This list is easily updated to include new companies/blogs.

\textit{Step 2 -- Search for keywords inside blogs using the Google Search API.} We adapted the keywords used in prior work~\citep{healthcare11202776} to suit the FM4SE and SE4FM contexts:

\smallskip
\noindent\fbox{%
\parbox{\dimexpr\linewidth-2\fboxsep-2\fboxrule}{%
\enquote{software} AND (\enquote{large language model} OR \enquote{large language models} OR \enquote{LLM} OR \enquote{LLMs} OR \enquote{foundation model} OR  \enquote{foundation models} OR \enquote{FM} OR \enquote{FMs} OR \enquote{generative AI} OR \enquote{GenAI}) site:\{url\}
}%
}
\smallskip

Using these keywords, we queried the Google Search API across all the blogs identified in Step 1, limiting the date range from August 10, 2023, to August 10, 2024. This process yielded 7,120 search results, including information about each result's URL, title, and snippet. We assigned a unique identifier (ID) to each search result, ranging from 0 to 7,120.
 
\textit{Step 3 -- Filter out non-blog posts and non-English results.}
The search results might contain unrelated pages such as index pages and author information pages. We filtered out URLs that contain strings such as \enquote{/index/} and \enquote{/author/} to eliminate non-blog posts. We used the \textit{langdetect} library\footnote{\url{https://github.com/Mimino666/langdetect}} to remove results where the title or snippet was not in English.

\textit{Step 4 -- Download the contents of the blog posts.} 
We downloaded all the content of all valid links. To further remove potential noise, we applied an outlier filter based on the interquartile range (IQR) to exclude content that was either too short or too long. After filtering, we retained 4,463 blog posts, each referenced by a unique ID from Step 2. Throughout this paper, individual blog posts are cited using their corresponding IDs (e.g., \blog{122} for blog post 122), allowing for direct reference to specific entries in the dataset. The full list of blog posts can be found in our online replication package~\cite{replication_package}.

\input{tables/llm_jury_golden_set}

\subsection{Identifying the FM4SE and SE4FM area, activities and discussed tasks in the blog posts}\label{subsec:classification}
We first label blog posts as FM4SE or SE4FM-related. Second, we label the FM4SE or SE4FM activities discussed in the blog posts. Finally, we summarize the activity-specific tasks that are discussed, to facilitate our manual review of the posts. Tables~\ref{tab:fmse_activity} and~\ref{tab:se4fm_activity} show the identified activities and tasks.
The first author, a PhD graduate with 5 years of SE research experience, conducted all labelling and validated through discussions with co-authors.

\textit{Step 1 -- Label blog posts as FM4SE or SE4FM with FM/LLM Jury.}
We used the FM/LLM Jury framework (Section~\ref{sec:llm_jury}) to classify blog posts as SE4FM, FM4SE, or unrelated. A golden dataset of 100 blog posts was randomly selected to evaluate this classification. 
Table~\ref{tab:llm_jury_golden_set} shows that the framework achieved an excellent agreement~($\kappa=0.92$) with human labels on the test set. We then applied the framework to the entire dataset of 4,463 blog posts, classifying 3,126 as unrelated, 156 as FM4SE and 1,122 as SE4FM.

\textit{Step 2 -- Label the FM4SE activities in the blog posts.}
We used the FM4SE activity labels from prior work~\citep{llm4se_survey_2024} and a golden set of 30 FM4SE posts was randomly sampled for evaluation. 
Following our iterative prompt construction process (Section~\ref{sec:llm_jury}), we reached an excellent agreement~($\kappa=0.91$) between the FM/LLM Jury and the human labels (see Table~\ref{tab:llm_jury_golden_set}). We applied the Jury to label all FM4SE activities in the 156 FM4SE blog posts. One blog post was labelled as `Other' as the FM/LLM Jury could not identify any specific FM4SE activities, ending up with 155 FM4SE blog posts in total.

\textit{Step 3 -- Label the SE4FM activities in the blog posts.}
Following a similar process, we labelled the SE4FM activities in the 1,122 SE4FM blogs. We created a list of SE4FM activity labels based on prior work~\citep{amershi_se4ml_2019, se4ai_survey_2022, hassan_aiware_2024}. We randomly sampled a golden set of 30 SE4FM blog posts and manually labelled the SE4FM activities. Following our prompt construction process, the FM/LLM Jury demonstrated strong agreement~($\kappa=0.86$) with the human labels. We used the best-performing prompt to label the SE4FM activities in all 1,122 SE4FM blog posts. 125 blog posts were labelled as `Other' as the FM/LLM Jury could not identify any SE4FM activities, ending up with 997 FM4SE blog posts in total.

\textit{Step 4 -- Identify the activity-specific tasks in the blog posts.} 
To facilitate our manual review of the blog posts, we prompted the FMs to take an additional step to identify the activity-specific tasks in the posts. 
For FM4SE blogs, we prompted the FMs to identify relevant tasks from a curated task set~\citep{llm4se_survey_2024, fan_llm4se_2023, liang_codeassist_2024}. Similarly, for SE4FM blogs, the FMs identified SE4FM-related tasks from a curated task set~\citep{amershi_se4ml_2019, se4ai_survey_2022, hassan_aiware_2024}. If no predefined tasks were relevant, we asked the FMs to generate new tags based on the blog content. In cases where the FMs proposed new task tags, we employed a human-in-the-loop process to determine whether to reclassify or ignore these tags. However, no human intervention was required for FM4SE posts, and eight new SE4FM task tags identified by the FM Gemini-1.5-Flash were resolved by the FM/LLM Jury.









%% file: tables/company_blog_sites.tex
\begin{table}[t]
\centering
\caption{Collected blogs with at least one relevant post.}
\label{tab:company_blog_sites}
\begin{tabular}{ll}
\toprule
Company  &  Blog \\
\midrule
AMD  &  \url{https://community.amd.com/t5/ai/} \\
Adobe  &  \url{https://blog.developer.adobe.com/} \\
      &  \url{https://blog.adobe.com/} \\
Alibaba  &  \url{https://www.alibabacloud.com/blog/} \\
Amazon  &  \url{https://www.amazon.science/blog} \\
       &  \url{https://aws.amazon.com/blogs/} \\
Cisco  &  \url{https://blogs.cisco.com/} \\
      &  \url{https://blog.talosintelligence.com/} \\
Google  &  \url{https://developers.googleblog.com/en/} \\
       &  \url{https://blog.google/} \\
       &  \url{https://cloud.google.com/blog/} \\
       &  \url{https://research.google/blog/} \\
       &  \url{https://deepmind.google/discover/blog/} \\
IBM  &  \url{https://research.ibm.com/blog} \\
    &  \url{https://developer.ibm.com/blogs/} \\
    &  \url{https://www.ibm.com/blog/} \\
Meta  &  \url{https://tech.facebook.com/engineering/} \\
     &  \url{https://research.facebook.com/blog/} \\
     &  \url{https://engineering.fb.com/} \\
Microsoft  &  \url{https://www.microsoft.com/en-us/research/blog/} \\
          &  \url{https://blogs.microsoft.com/} \\
          &  \url{https://devblogs.microsoft.com/} \\
          &  \url{https://techcommunity.microsoft.com/t5/} \\
Nvidia  &  \url{https://blogs.nvidia.com/} \\
       &  \url{https://developer.nvidia.com/blog/} \\
Oracle  &  \url{https://blogs.oracle.com/} \\
Qualcomm  &  \url{https://www.qualcomm.com/developer/blog} \\
SAP  &  \url{https://community.sap.com/t5/technology-blogs-by-sap} \\
Salesforce  &  \url{https://www.salesforce.com/blog/} \\
           &  \url{https://engineering.salesforce.com/blog/} \\
           &  \url{https://blog.salesforceairesearch.com/} \\
\bottomrule
\end{tabular}
\end{table}

%% file: tables/llm_jury_golden_set.tex
\begin{table}[t]
\centering
\caption{Cohen's Kappa between LLMs and human labels on the golden set for SE-FM area, FM4SE activity, and SE4FM activity classification tasks.}
\label{tab:llm_jury_golden_set}
\begin{tabular}{lrrr}
\toprule
Model                  & \begin{tabular}[c]{@{}r@{}}SE-FM\\ Area\end{tabular} & \begin{tabular}[c]{@{}r@{}}FM4SE\\ Activities\end{tabular} & \begin{tabular}[c]{@{}r@{}}SE4FM\\ Activities\end{tabular} \\ \midrule
Gemini-1.5-Flash-002   & 0.64                                                 & 0.81                                                       & 0.69                                                       \\
GPT-4o-mini-2024-07-18 & 0.70                                                 & 0.81                                                       & 0.77                                                       \\
Qwen2-72B-Instruct     & 0.85                                                 & 0.76                                                       & 0.73                                                       \\ \midrule
\textbf{FM/LLM Jury}        & \textbf{0.92}                                        & \textbf{0.91}                                              & \textbf{0.86}                                              \\ \bottomrule
\end{tabular}
\end{table}

%% file: sections/results.tex

\input{sections/rq1}
\input{sections/rq2}

%% file: sections/rq1.tex
\section{RQ1: \rqone}\label{sec:rq1}

\motivation
Industry practitioners are at the forefront of applying FMs to SE, sharing practical insights and real-world experiences through blogs. While academic research has explored many aspects of FM4SE, the industry's perspective remains underexplored. This study analyzes industry blogs to uncover key FM4SE activities and tasks discussed by practitioners, providing insights into real-world applications of FMs in SE.

\approach
We used the FM/LLM Jury to label the FM4SE activities and tasks in the 155 FM4SE blog posts. 
To avoid overrepresentation, we counted each activity and task uniquely per company, even if they were mentioned in multiple blog posts from the same company. To gain insights into how these activities and tasks are discussed, we manually reviewed the selected blog posts. We also compared our findings with those reported by \citet{llm4se_survey_2024}.

\input{tables/fm4se}


\subsection{Software development}

\textbf{Although code generation is the most prominent task, FMs are used for many other tasks in the software development process.} As shown in Table~\ref{tab:fmse_activity}, code generation emerges as the most prominent task. Practitioners report leveraging FMs to generate code in modern languages~\blog{B140} such as Python and Java, but there is also growing attention for legacy systems, with FMs being used to generate COBOL code~\blog{B830}. Additionally, FMs are applied to specialized domains such as SQL query generation~\blog{B183}, and domain-specific languages (DSLs) tailored to industry-specific needs such as semiconductor design~\blog{B8}. The flexibility of FMs to adapt across various programming languages and domains highlights their versatility in software development.

A topic that is closely related to, and often discussed together with code generation is \taskfmt{code completion}. A notable technique is \textit{fill-in-the-middle}~\citep{fried2023incoder} for code completion, which allows models to complete code based on partial inputs~\blog{B157}, further expanding the usability of FMs in practical coding environments.

\textbf{Beyond code generation and completion, FMs are increasingly being integrated into software development as \taskfmt{code assistants}, offering a range of functionalities.} These assistants not only generate and complete code but also help with \taskfmt{code understanding}, \taskfmt{code summarization}, \taskfmt{code optimization}, and \taskfmt{API recommendations}. For instance, code assistants can understand code and explain it to developers~\blog{B5668}, summarize code changes for reviews~\blog{B322}, optimize code for performance~\blog{B3357}, or recommend APIs based on both public and private code repositories~\blog{B3177}. This multifaceted support streamlines the development process, boosting developer productivity and efficiency.

\subsection{Software quality assurance}

\textbf{Vulnerability detection is the most frequently discussed software quality assurance (QA) task.} Practitioners employ FMs to automate \taskfmt{common vulnerabilities and exposures~(CVE) detection} and analysis~\blog{B101}. Other tasks under this category include \taskfmt{test generation and automation}, where FMs are used to generate test cases based on the functionality of a given code. For instance, the FM can suggest test cases for invalid inputs, edge cases, and error handling~\blog{B5518}. For \taskfmt{debugging} tasks, FMs can suggest where to insert logging and exception handling to track code execution and errors, and FMs can also be used to detect anomalies and fix common issues such as syntax and logical errors~\blog{B5465}.

\subsection{Software maintenance}

\textbf{The use of FMs in software maintenance focuses on the refactoring, translation, and transformation of existing codebases.} Practitioners often discuss these tasks in the context of modernizing legacy systems. For instance, FMs are employed to \taskfmt{refactor} and \taskfmt{translate} legacy COBOL code into Java~\blog{B769}. In addition, upgrading Java applications to newer Long-Term Support (LTS) versions~\blog{B369} or migrating Java codebases to cloud-based infrastructures~\blog{B6378} are tasks frequently associated with \taskfmt{code transformation}. Using FMs for these tasks helps industries transition from older systems to modern architectures more efficiently.

\subsection{Software management, requirement engineering, and software design}

\textbf{Software management receives the least attention in FM4SE blogs, and no discussions were found regarding requirements engineering or software design.} Practitioners use FMs to generate \taskfmt{software tool configurations} for managing cloud infrastructure components~\blog{B5625}. We did not find discussions on \taskfmt{requirements engineering} or \taskfmt{software design} in FM4SE blogs. One blog, which was initially misclassified as covering requirements analysis~\blog{B1182}, was actually about using FMs to extract developer intent from code comments or function documentation for generating formal postconditions. Overall, FM-based requirements engineering and software design remain an underreported area in the industry.

\subsection{Comparison with Academic Research}

Software development is the most discussed activity in both industry blog posts and SE research papers~\citep{llm4se_survey_2024}. Both practitioners and researchers frequently highlight code generation as a key task. However, there are some notable differences. Regarding software maintenance, for example, industry blogs focus more on code refactoring and revision, while academic research papers focus more on program repair. Additionally, while \citet{llm4se_survey_2024} reported that 4.3\% of surveyed papers (17 out of 395) cover requirements engineering, we found no substantial discussion on this topic in the industry blogs we analyzed~(except the one misclassified).

\hypobox{
\textbf{RQ1 Summary:} 
Software development tasks, particularly code generation, are the most frequently discussed across FM4SE blogs. FMs are increasingly integrated as code assistants, providing developers with multifunctional tools to boost productivity. In the domain of software quality assurance, vulnerability detection is the dominant task, while software maintenance activities are primarily focused on refactoring and transforming existing codebases. Software management, requirements engineering, and software design receive less attention in these industry blogs.
}



%% file: tables/fm4se.tex
\begin{table}[t]
\centering
\caption{FM4SE activities and tasks that are discussed in industry blog posts. The Comp. column indicates the number of companies publishing blog posts about the task.}
\label{tab:fmse_activity}
\begin{tabular}{lrlr}
\toprule
Activity & Posts & Task & Comp. \\ \midrule
\multirow{7}{*}{\begin{tabular}[c]{@{}l@{}}Software\\ development\end{tabular}} & 101 & Code generation & 11 \\
 & 54 & Code completion & 8 \\
 & 12 & Code assistant & 5 \\
 & 10 & Code understanding & 5 \\
 & 6 & Code summarization & 4 \\
 & 3 & Code optimization & 3 \\
 & 1 & API recommendation & 1 \\ \midrule
\# Total & 121 &  & 11 \\ \midrule
\multirow{3}{*}{\begin{tabular}[c]{@{}l@{}}Software\\ quality\\ assurance\end{tabular}} & 10 & Vulnerability detection & 3 \\
 & 5 & Debugging & 3 \\
 & 4 & Test generation/automation & 2 \\ \midrule
\# Total & 17 &  & 5 \\ \midrule
\multirow{6}{*}{\begin{tabular}[c]{@{}l@{}}Software \\ maintenance\end{tabular}} & 4 & Code refactoring or revision & 3 \\
 & 7 & Code transformation & 2 \\
 & 3 & Code translation & 2 \\
 & 2 & Code review & 1 \\
 & 1 & Program repair & 1 \\
 & 1 & Software operations & 1 \\
 & 1 & Log analysis & 1 \\ \midrule
\# Total & 15 &  & 4 \\ \midrule
\begin{tabular}[c]{@{}l@{}}Software\\ management\end{tabular} & 1 & Software tool configuration & 1 \\ \midrule
\begin{tabular}[c]{@{}l@{}}Requirement\\ engineering\end{tabular} & 1 & Requirements analysis & 1 \\ \midrule
\begin{tabular}[c]{@{}l@{}}Software design\end{tabular} & 0 & -- & 0 \\ \midrule
\# Total & 155 &  & 11 \\ \bottomrule
\end{tabular}
\end{table}

%% file: sections/rq2.tex
\section{RQ2: \rqtwo}\label{sec:rq2}

\motivation
As FMs are integrated into production systems, applying SE principles to the development cycle of FM-based systems becomes increasingly important. While academic research studied SE for AI-based systems~\citep{se4ai_survey_2022}, FM-based systems present unique challenges~\citep{hassan_aiware_2024}, such as the resource-intensive nature of FMs, and the complexities involved in fine-tuning, deployment, and monitoring at scale. Industry blog posts offer valuable insights into how practitioners adapt SE principles for developing, deploying, managing, and scaling FMs in real-world settings. This study gives an overview of key SE4FM activities and tasks discussed in these blog posts.

\approach
We used the FM/LLM Jury to label the SE4FM activities and tasks in the 997 SE4FM blog posts. 
Similar to Section~\ref{sec:rq1}, we counted each activity and task uniquely per company and manually reviewed selected blog posts for each task. We did not compare the discussion frequency with previous research like we did in Section~\ref{sec:rq1}, because there exists no prior survey on SE4FM, and SE4FM activities and tasks are quite different from those for SE4ML~\cite{se4ai_survey_2022}.


\subsection{Model deployment \& operation}

\textbf{Model deployment \& operation is the most frequently discussed activity in SE4FM industry blog posts.} \taskfmt{Model deployment on cloud} is the dominant task (see Table~\ref{tab:se4fm_activity}), reflecting the industry's reliance on cloud environments for hosting foundation models. Foundation models are very resource intensive. For example, a large model such as Meta Llama 3 (with 405 billion parameters~\citep{dubey2024llama3herdmodels})~\blog{B1749,B5439,B6691} requires $\sim$810GB of GPU VRAM for inference, $\sim$3.25TB for fine-tuning~\cite{hf_llama3} and much more for training from scratch. The cloud facilitates using such large models on-demand without the need for buying very expensive hardware. 

For \taskfmt{model serving \& scaling}~\blog{B111,B627,B6028}, techniques such as speculative decoding~\cite{chen2023acceleratinglargelanguagemodel} accelerate model inference by using draft models for faster response times~\blog{B5308}. In addition, automatic model scaling ensures that resources automatically adjust to workload needs~\blog{B701,B4518}. \taskfmt{Model monitoring}~\blog{B6659} is another key task, involving tracking token usage~\blog{B406,B920} and monitoring system metrics such as memory and GPU load~\blog{B5192}.

\textbf{While cloud deployment dominates, there is increasing interest in \taskfmt{model deployment on local} devices such as edge or mobile devices, and PCs.} For example, practitioners deploy FM-based medical chatbots for healthcare applications on edge devices, to facilitate data privacy~\blog{B139}. Another reason to deploy FMs on smaller devices is to overcome the GPU supply-and-demand problem, which makes it hard for many companies to integrate FMs into their products~\blog{B1355}. 
To enable running the resource-intensive FMs on relatively small devices, companies frequently use \taskfmt{model compression} techniques~\citep{zhu_modelcompress_2024} to reduce the required resources~\blog{B1744,B1993,B5463,B5714}. For example, quantization techniques (e.g., 4-bit precision) enable running models on CPUs, avoiding the need for GPUs~\blog{B669}. Compression techniques are not limited to text models: model quantization is also applied to image generative models such as Stable Diffusion~\cite{rombach2021highresolution}~\blog{B2254}. Several libraries are used by practitioners that assist with running FMs on CPU, such as  \textit{LLaMA.cpp}\footnote{\url{https://github.com/ggerganov/llama.cpp}} and  \textit{ExLlama}\footnote{\url{https://github.com/turboderp/exllama}}~\blog{B1355}. Also, several practitioners describe how the use of Neural Processing Units (NPUs) can facilitate running FMs locally~\blog{B914,B1750}.

\input{tables/se4fm}




\subsection{System architecture \& orchestration}

\textbf{System architecture \& orchestration activities, including building AI agents and model \& prompt chaining, have become a popular topic in SE4FM industry blog posts.} One of the most frequently discussed tasks is \taskfmt{model \& prompt chaining}, which is used to manage complex workflows by breaking them into smaller, more manageable steps, each handled by different models or prompts~\blog{B881}. For example, a task such as responding to customer reviews might be divided into steps like filtering harmful content, performing sentiment analysis, and generating an appropriate response~\blog{B4881}. Tools and frameworks like \textit{LangChain}~\citep{Chase_LangChain_2022} and \textit{PromptFlow}~\citep{promptflow} are commonly mentioned as practical solutions for implementing these chained workflows~\blog{B1357}.

A closely related discussion topic is \taskfmt{building AI agents}, which extend the functionality of FMs by integrating external tools and \taskfmt{orchestrating workflows}.
In multi-agent systems, multiple AI agents collaborate using a complex workflow to handle different parts of a task~\blog{B1166}.
These AI agents can autonomously decide which tools to use, retrieve necessary data, and execute predefined plans based on user input or real-time data~\blog{B414}. 
Another key feature of AI agents is their ability to leverage working memory, allowing them to retain information from previous interactions or external tool outputs, which can be critical for managing long-term tasks~\blog{B442}. 

Enterprise \taskfmt{development platforms \& studios} provide support for building FM-based systems based on chaining or AI agents~\blog{B414,B1166,B4881}. These platforms, such as \textit{FMArts}~\citep{hassan_aiware_2024}, simplify the orchestration of workflows~\blog{B358}, allowing developers to easily integrate FMs with external systems. These platforms also support \taskfmt{implementing guardrails}. Guardrails can take the form of filters that are applied to user inputs or LLM outputs~\blog{B4677}, or can be embedded in the prompts to guide the model's responses~\blog{B2456}. 




\subsection{Data management}

\textbf{Data management has evolved in the FM era, with new techniques supporting the vast amounts of both structured and unstructured data.} At the core of this evolution is the shift to specialized, more dynamic data management techniques. The most frequently discussed task is \taskfmt{Retrieval-Augmented Generation (RAG)}~\citep{lewis_rag_2020}, which combines the use of private datasets (i.e., data on which the FM was not trained before such as proprietary data) with FMs. With RAG, the FM generates its response based on the prompt and information in the private dataset~\blog{B4270}. Several practitioners discuss how they build datasets for RAG~\blog{B6492,B6688} using techniques including document chunking, embedding, and vector storage. 
Another example of usage of RAG by practitioners is GraphRAG~\citep{edge2024localglobalgraphrag} which enhances RAG by generating knowledge graphs from the private data which can then be used for prompt augmentation~\blog{B1155,B1156}. 

\textbf{There is a great amount of discussion of \taskfmt{specialized databases}, particularly vector databases.} Specialized databases are key to enabling RAG. These databases support semantic search by indexing unstructured data like text and images, making FM-based retrieval faster and more accurate~\blog{B1571}. Advanced features include multimodal search, where users retrieve image or video content using text queries~\blog{B5035}. This shift from traditional keyword-based search to semantic search also integrates with SQL queries for managing both structured and unstructured data~\blog{B1426}. As data management moves beyond traditional data types (e.g., rows, columns, JSON), vector-based storage and retrieval systems are becoming more important~\blog{B5370}. Likewise, \taskfmt{embedding as feature engineering} is becoming increasingly important in FM-based systems, enabling text, images, and structured data to be converted into numerical vectors that FMs can process~\blog{B1982}. Multimodal embeddings, which map both text and images into a shared vector space, are particularly useful in cross-modal applications such as text-to-image search or video retrieval~\blog{B5815}.

\textbf{Synthetic data generation provides scalable, domain-specific data without privacy risks, reducing reliance on real-world datasets.}
As FM data requirements grow, synthetic data is increasingly used to address the challenges of high-quality \taskfmt{data collection}~\blog{B35,B4210}. 
In parallel, \taskfmt{automated data labelling} is being transformed by model-assisted approaches. For example, the Recognize Anything Model~(RAM)~\cite{zhang2023recognizeanythingstrongimage} can automatically label visual datasets, enabling users to search for images or videos using natural language queries~\blog{B3649}. Additionally, human-in-the-loop is applied for combining model-generated annotations with manual oversight to ensure accuracy while reducing the time and cost associated with traditional labelling methods~\blog{B5181}.


\textbf{There is a noticeable push on data privacy in \taskfmt{data cleaning \& preparation}.} \taskfmt{Data cleaning \& preparation} remains essential, but data privacy has become an important focus. Since the data used for customizing FMs could contain Personally Identifiable Information~(PII), data anonymization techniques such as differential privacy~\blog{B5934} are used to remove PII~\blog{B1286,B1309,B3213}. Beside data privacy, removing duplicate is frequently applied to preprocess the dataset to ensure data quality~\blog{B188,B804,B2326}.

\subsection{Model customization}

\textbf{Model customization is achieved through fine-tuning techniques for adapting FMs to specific application needs.} \taskfmt{Fine-tuning methods} such as supervised fine-tuning~(SFT) tune the entire model on domain-specific data, rather than train it from scratch. Enterprise platforms now support no-code fine-tuning, simplifying the process and accelerating development~\blog{B4777}. Open source libraries such as Hugging Face’s PEFT~\citep{peft} support both full fine-tuning and \taskfmt{Low-Rank Adaptation (LoRA)}~\blog{B353}. LoRA is an efficient approach where original model parameters are frozen and injected with trainable matrices. LoRA reduces the number of trainable parameters and lowers GPU requirements, making it cost-effective~\blog{B109}. With different LoRA adapters, a single FM can adapt to handle different tasks. Platforms support dynamic loading and caching of LoRA adapters, offering flexibility and optimizing performance~\blog{B2462}. \taskfmt{RLHF} is used to align models with user preferences to improve their experience~\blog{B656} and can also be applied to image models~\blog{B2367}.


\subsection{Evaluation \& quality assurance}

\textbf{SE4FM blog posts outline practical strategies for ensuring the safety, fairness, and trustworthiness of FMs through systematic evaluation \& QA processes.} 
With the diverse applications of FMs, establishing robust \taskfmt{model evaluation} frameworks is essential to ensure models meet operational requirements~\blog{B898}. For example, for FM-based systems with RAG integration, an evaluation framework includes metrics such as answer relevance, context precision, and recall to assess the effectiveness of model outputs~\blog{B1305}. Ensuring \taskfmt{model safety \& compliance} is particularly critical in high-stakes industries. Industry blogs highlight the use of adversarial testing to identify model vulnerabilities, while automated raters are often deployed to perform consistent safety assessments~\blog{B437}. Standardized benchmarks are also leveraged to ensure models meet security and compliance standards, especially in regulated industries~\blog{B845}.

Practitioners are increasingly adopting \taskfmt{automated testing strategies for FMs}, which often use academic benchmarks like BIG-bench~\blog{B158}. However, custom datasets tailored to specific domain requirements are also vital for evaluating FMs in domain-specific applications~\blog{B845}. One emerging approach in this area is the use of LLM-as-a-judge techniques, which leverage LLMs to provide scalable and consistent evaluations~\blog{B158}. Additionally, adversarial testing is used to strengthen models against potential threats by uncovering weaknesses that may not surface under traditional testing methods~\blog{B845}. Another emerging task is \taskfmt{model explainability \& interpretability}, which is important particularly in sensitive industries. Tools that generate natural language explanations for model outputs are becoming common, helping developers understand why certain test cases pass or fail~\blog{B4776}. This increases transparency and aligns FMs with best practices for software engineering.

Some practitioners discuss \taskfmt{model fairness \& bias}, as biased outputs from FMs can lead to ethical concerns or operational risks. To reduce bias in FM-generated outputs, techniques such as prompt engineering and scenario testing are employed~\blog{B1295}. Adversarial testing and diverse rater systems are also leveraged to ensure fairness and prevent harmful outputs~\blog{B437}. Additionally, practitioners highlight the importance of human oversight in critical decision-making processes to mitigate risks associated with harmful or biased outputs~\blog{B1064}. To enhance trust in FM-based systems, practitioners conclude and follow a set of best practices for AI security, such as the Secure AI Framework (SAIF)~\blog{B697}.

\subsection{Prompt construction}

\textbf{Prompt construction receives the least attention, and no discussions were found regarding requirements engineering.} \taskfmt{Prompt engineering} techniques are discussed in the blog posts, such as structured prompts~\blog{B374} and multi-shot prompts~\blog{B4401}. In database contexts, prompts consider schema, query history, and user-specific factors to generate SQL queries~\blog{B3601}. In addition, \taskfmt{automated prompt generation} techniques are explored through dynamic metaprompts which are optimized for greater control and adaptability~\blog{B1172}. For tasks like text-to-image generation, prompts are improved based on semantic search and user context~\blog{B3754}. In addition, prompt compression is proposed to automatically reduce prompt length without sacrificing essential information~\blog{B1173}.

\hypobox{
\textbf{RQ2 Summary:}
The most discussed activities in SE4FM blog posts are model deployment \& operation, with a focus on cloud hosting and model serving \& scaling. With regards to system architecture, trends include prompt chaining, workflow orchestration for FMs, and AI agents. Data management emphasizes RAG and vector databases to support unstructured data and information retrieval. Model customization relies on fine-tuning methods such as full fine-tuning, LoRA, and RLHF. Evaluation \& quality assurance practices focus on automated testing, safety, trustworthiness, and bias mitigation, while prompt engineering receives limited attention. Discussion about requirements engineering is not found in SE4FM blog posts.
}





%% file: tables/se4fm.tex
\begin{table}[t]
\caption{SE4FM activities and tasks that are discussed in industry blog posts. The Comp. column indicates the number of companies publishing blog posts about the task.}
\centering
\label{tab:se4fm_activity}
\begin{tabular}{lrlr}
\toprule
Activity & Posts & Task & Comp. \\ \midrule
\multirow{5}{*}{\begin{tabular}[c]{@{}l@{}}Model \\ deployment \\ \& operation\end{tabular}} & 237 & Model deployment on cloud & 13 \\
 & 168 & Model serving \& scaling & 12 \\
 & 30 & Model monitoring & 8 \\
 & 31 & Model compression & 7 \\
 & 39 & Model deployment on local & 6 \\ \midrule
\# Total & 373 &  & 13 \\ \midrule
\multirow{5}{*}{\begin{tabular}[c]{@{}l@{}}System \\ architecture \\ \& orchestration\end{tabular}} & 94 & Model \& prompt chaining & 11 \\
 & 108 & Workflow orchestration & 10 \\
 & 83 & Building AI agents & 10 \\
 & 99 & Development platform \& studio & 9 \\
 & 6 & Implementing guardrails & 3 \\ \midrule
\# Total & 287 &  & 12 \\ \midrule
\multirow{6}{*}{\begin{tabular}[c]{@{}l@{}}Data \\ management\end{tabular}} & 90 & RAG integration & 11 \\
 & 48 & Specialized databases & 8 \\
 & 40 & Dataset cleaning \& preparation & 8 \\
 & 43 & Dataset collection & 6 \\
 & 7 & Feature engineering & 5 \\
 & 3 & Dataset labelling & 2 \\ \midrule
\# Total & 182 &  & 11 \\ \midrule
\multirow{3}{*}{\begin{tabular}[c]{@{}l@{}}Model \\ customization\end{tabular}} & 85 & General fine-tuning & 10 \\
 & 20 & LoRA & 4 \\
 & 10 & RLHF & 4 \\ \midrule
\# Total & 104 &  & 11 \\ \midrule
\multirow{6}{*}{\begin{tabular}[c]{@{}l@{}}Evaluation \\ \& quality \\ assurance\end{tabular}} & 17 & Model evaluation & 6 \\
 & 15 & Model safety \& compliance & 6 \\
 & 6 & Model risk \& trust & 4 \\
 & 5 & Testing strategies & 4 \\
 & 2 & Model fairness \& bias & 2 \\
 & 1 & Model explainability & 1 \\ \midrule
\# Total & 40 &  & 7 \\ \midrule
\multirow{2}{*}{\begin{tabular}[c]{@{}l@{}}Prompt \\ construction\end{tabular}} & 10 & Prompt engineering & 5 \\
 & 4 & Automated prompt generation & 3 \\ \midrule
\# Total & 11 &  & 6 \\ \midrule
\begin{tabular}[c]{@{}l@{}}Requirements \\ engineering\end{tabular} & 0 & -- & 0 \\ \midrule
\# Total & 997 &  & 14 \\ \bottomrule
\end{tabular}
\end{table}

%% file: sections/discussion.tex
\section{Discussion of Future Research Directions}\label{sec:discussion}




\subsection{Research Directions for FM4SE}

\textbf{Research Direction 1: Using FMs for modernization and transformation of legacy code.} Practitioners applied FMs to translate legacy systems into modern languages such as Java~\blog{B769}, upgrade to newer language versions~\blog{B369}, and migrate to cloud-based infrastructures~\blog{B6378}. This process mostly relies on FMs for code translation, however, \citet{pan_codetrans_2024} highlight challenges in applying FMs to translate code in real-world projects. To address these challenges, researchers should explore methodologies that enhance FM performance for automating complex system migrations, including the transformation of entire legacy codebases and architectures.

\textbf{Research Direction 2: Evaluating code assistants in software development workflows for tasks other than code generation.} \citet{liang_codeassist_2024} conducted a survey on the usability of code assistants focusing on the code generation task. However, industry discussions highlight that code assistants are employed for other tasks such as code understanding~\blog{B5668}, code summarization~\blog{B3357}, and API recommendation~\blog{B3177} as well. This broader integration, including their potential role as AI teammates~\citep{hassan_se3_2024}, suggests opportunities for researchers to expand studies on code assistants beyond code generation. 

\textbf{Research Direction 3: Real-world validation of research on applying FMs in software management, requirements engineering, and software design.} Our analysis of industry blog posts shows that discussions around FM applications in these activities are rare~(see Table~\ref{tab:fmse_activity}). Likewise, though software engineering researchers have explored using FMs for requirements engineering tasks such as requirements classification and traceability automation, such studies remain relatively rare as well~\citep{llm4se_survey_2024}. Considering the capabilities of FMs in handling natural languages and programming languages, they should be well-suited for these tasks, hence the lower number of (reported) applications in this area is surprising. Researchers should aim to bridge the gap by identifying the barriers to and developing tools that apply FM techniques in these underreported activities. Such efforts could help bring research advancements in these areas closer to practical industry applications, showcasing the value of FM4SE for a broader range of software engineering activities.

\subsection{Research Directions for SE4FM}

\textbf{Research Direction 4: Researchers should expand research on SE4FM.} Our findings in Sections~\ref{sec:rq1} and \ref{sec:rq2} show that practitioners discussed SE4FM much more often than FM4SE, with 997 blog posts focused on SE4FM compared to 155 blog posts on FM4SE. Activities such as model deployment \& operation, system architecture, data management, and model customization are frequently mentioned by more than half of the companies~(Table~\ref{tab:se4fm_activity}). 
Although the lower number of posts on FM4SE does not necessarily indicate a lack of interest in these topics, there seems to be a disconnect between academic research and practitioner activities.
The results of our study highlight the growing importance of adapting SE practices to support the development lifecycle of FM-based systems. Researchers should explore these areas to bridge the gap between industry practices and academic research, thereby supporting the evolving needs of FM-based systems.

\textbf{Research Direction 5: Performance engineering for FM-based systems.} Performance optimization of FM-based systems~\citep{hassan_aiware_2024} is a critical area that practitioners frequently discussed (in several tasks). Techniques such as model compression and model serving \& scaling are employed to reduce the computational resources needed for deploying FMs~\blog{B701,B5463}. Researchers could contribute by formalizing and standardizing such techniques, as well as exploring new methods to reduce the computational overhead of fine-tuning and deploying FMs. The resource-intensive nature of FMs introduces challenges related to managing large-scale parallelism and ensuring memory efficiency, especially for models with billions of parameters~\blog{B5192}. Extending traditional load testing methods~\citep{jiang_loadtest_2015} to FM inference pipelines could help ensure that these models scale effectively and meet the performance requirements of real-time applications.

\textbf{Research Direction 6: Investigate the impact of the shift from full model training to fine-tuning on software engineering activities such as dependency and asset management.} Industry blog posts highlight the increasing reliance on libraries for model fine-tuning and model inference, such as PEFT from Hugging Face~\blog{B353} and LLaMA.cpp~\blog{B1355}. Researchers should investigate how these emerging tools and techniques integrate into existing SE workflows and assess their impact on model quality, performance, and usability. For example, the reduced set of trainable parameters used by LoRA can be stored separately as an adapter~\blog{B2462}, which introduces new challenges in asset management. Researchers should investigate how to manage and optimize these adapters within the context of FM-based systems.

\textbf{Research Direction 7: Supporting FM workflow orchestration and AI agents through software engineering activities.} Practitioners are integrating FMs not as standalone components but as part of complex workflows that chain together multiple models, prompts, and external APIs~\blog{B881,B1357}. Furthermore, AI agents that autonomously manage tasks and orchestrate tools are becoming prevalent in industry~\blog{B414,B1166}. Researchers should explore design patterns, best practices, and frameworks that support the development of these FM-based systems.

\textbf{Research Direction 8: Supporting the evolving data pipelines in FM-based systems.}  While traditional data collection, cleaning, and labelling methods  remain essential, industry practices are increasingly incorporating synthetic data generation and automated labelling using FMs~\blog{B4210,B5181}. Synthetic data generation is used to produce large volumes of domain-specific data without privacy concerns~\blog{B35}. However, as reliance on synthetic and automatically labeled data increases, researchers must investigate the trade-offs between data quality, model performance, and ethical considerations. Human-in-the-loop approaches, where models assist with but do not completely replace manual labelling, also offer a promising area for further exploration~\blog{B5181}. Additionally, the growing use of specialized databases, such as vector databases for efficient retrieval of unstructured data~\blog{B1571}, demands research into how these evolving data management strategies affect the development lifecycle of FM-based systems, particularly when integrating multimodal data~\blog{B5035}.

%% file: sections/threats.tex
\section{Threats to Validity}\label{sec:threats}


\textit{Internal Validity.} Industry blogs may not fully represent a company’s official or comprehensive views, as posts are often authored by individuals or specific teams. This introduces a risk of bias since not all companies maintain blogs which could potentially skew our dataset. In addition, there could be marketing trends or recent product launches, which could further skew the dataset. To mitigate this threat, we not only report the number of blogs mentioning a specific activity, but also the number of companies discussing those activities in our analysis. Additionally, we realize that any overview of discussed topics in fast moving areas such as FM4SE and SE4FM is only a snapshot of the situation at that time, which can quickly change. We want to emphasize that our FM-powered surveying approach is automated and flexible, and can be rerun easily for different companies, blog posts and possibly even different areas. 

The prompt optimization approach in Section~\ref{sec:methodology} is currently manual. In future work, we plan to explore automated prompt optimization frameworks, such as \texttt{dspy}~\cite{khattab2023dspy, khattab2022demonstrate}.

\textit{External Validity.} The FM/LLM Jury comprises models of different sizes. If using smaller models, they might lack emergent abilities~\citep{wei2022emergentabilitieslargelanguage} found in larger ones which could affect the accuracy of the labels. Furthermore, our findings are based on blogs from large companies and may not generalize to smaller ones which are limited by resource constraints.

%% file: sections/conclusion.tex
\section{Conclusion}\label{sec:conclusion}

This study bridges the gap between academic research and industry practices by analyzing 1,152 industry blog posts related to FM4SE and SE4FM from leading technology companies. We used an FM/LLM Jury of multiple FMs to automatically label and summarize the contents. We uncovered key activities and trends discussed by industry practitioners in the intersection of SE and FMs.
Our main findings are:
\begin{itemize}
    \item Although code generation is the most prominently discussed task in FM4SE blog posts, FMs are used for many other tasks in the software development process, including code completion, code understanding, and code summarization. Software engineering researchers should investigate how to leverage FMs to support an even broader range of software engineering activities. 
    \item In SE4FM blog posts, model deployment \& operation and system architecture \& orchestration are the most frequently discussed activities. Software engineering researchers should investigate how these activities can be further supported through software engineering. 
\end{itemize}

Our findings offer valuable insights into how industry leaders are leveraging FMs and SE. We provide researchers with eight promising research directions to explore. Additionally, we demonstrate the potential of using an FM-powered surveying approach for automating grey literature surveys in rapidly evolving fields like SE and ML.